\documentclass[twocolumn,showpacs,preprintnumbers,amsmath,amssymb]{revtex4}

\usepackage{graphicx}
\usepackage{dcolumn}
\usepackage{bm}

\newcommand{\bq}{\begin{equation}}
\newcommand{\eq}{\end{equation}}
\newcommand{\bqa}{\begin{eqnarray}}
\newcommand{\eqa}{\end{eqnarray}}
\newcommand{\nn}{\nonumber \\}

\def\be     {\begin{equation}}
\def\ee     {\end{equation}}
\def\bea        {\begin{eqnarray}}
\def\eea        {\end{eqnarray}}
\def\bnn    {\begin{eqnarray*}}
\def\enn    {\end{eqnarray*}}

\begin{document}

\title{Role of non-magnetic disorder in a doped U(1) spin liquid}
\author{Ki-Seok Kim}
\affiliation{Korea Institute of Advanced Study, Seoul 130-012,
Korea}
\date{\today}

\begin{abstract}
Recently we investigated a role of non-magnetic disorder on the
stability of a $U(1)$ spin liquid ($U1SL$) [cond-mat/0407151;
Phys. Rev. B (R) accepted]. In the present paper we examine an
effect of the non-magnetic disorder on a doped $U1SL$. In a recent
study [cond-mat/0408236] we have shown that the doped $U1SL$ shows
deconfined massive spinon excitations in the superconducting phase
as a result of holon condensation. We find that the massive spinon
is trapped by the non-magnetic disorder. Owing to the localized
spin the non-magnetic disorder acts as magnetic one.
\end{abstract}


\maketitle

It is now believed that non-magnetic disorder induces a local
magnetic moment of $S = \frac{1}{2}$ in underdoped
cuprates\cite{Theory_1,Theory_1_1,Theory_2,Theory_3,Theory_4}. The
localized moment is expected to result in a Curie-Weiss behavior
of a magnetic susceptibility\cite{Theory_1,Theory_2,Theory_4} and
increase antiferromagnetic correlations\cite{Theory_1_1}. Despite
of these successful descriptions there exists controversy. Sachdev
et al. claimed that emergence of the local moment in a
non-magnetic impurity can be understood by confinement of a spinon
in the underdoped cuprates\cite{Theory_3,Confinement}. But the
previous studies\cite{Theory_1,Theory_2} are based on a
deconfinement phase which is described by a slave particle mean
field theory. It has been correctly pointed out that in the slave
boson mean field theories strong gauge fluctuations mediating
interactions between the slave particles are not appropriately
treated\cite{Nayak,DHLee}. Thus it is not clear that if strong
gauge fluctuations are appropriately treated, the conclusions in
the previous studies\cite{Theory_1,Theory_2} are sustained. It is
necessary to treat both the gauge fluctuations and non-magnetic
impurities on equal footing.

Lastly Senthil et al. claimed that instanton excitations
originating from strong gauge fluctuations can be suppressed by
critical fluctuations of massless Dirac fermions in the large
flavor limit\cite{U1SL}. As a result an effective field theory to
describe a spin degree of freedom in the underdoped cuprates is
obtained to be $QED_3$ in terms of massless Dirac fermions
(spinons carrying only spin $\frac{1}{2}$) interacting via
non-compact U(1) gauge fields\cite{U1SL,Senthil_Lee}. The state
described by this critical field theory is usually called a U(1)
spin liquid ($U1SL$).

Recently we examined a role of non-magnetic disorder on the
stability of the $U1SL$ using a renormalization group
calculation\cite{Kim_U1SL}. In other words, we treated both strong
gauge fluctuations and non-magnetic impurities on equal footing.
In the study we found that the $U1SL$ remains stable against weak
disorder. The IR fixed point of $QED_3$ to govern the $U1SL$ is
stable against the presence of the weak non-magnetic disorder and
thus the $U1SL$ is sustained. This shows that the massless Dirac
spinon is not localized by the weak disorder in the $U1SL$, which
is in contrast with the results of the slave particle mean field
theories\cite{Theory_1,Theory_2}.

In the present paper we investigate an effect of the non-magnetic
disorder on a hole doped $U1SL$. Not only strong gauge
fluctuations and non-magnetic impurities but also hole doping is
treated on equal footing. Recently we showed that hole doping to
the $U1SL$ can give rise to serious change in the
$U1SL$\cite{Kim_SU2}. In the study we claimed that condensation of
holons (representing doped holes) results in a fermion (spinon)
zero mode in the presence of an instanton potential. As a result
instanton excitations are suppressed and thus deconfinement of the
spinons and holons are expected to occur. Further, in the
superconducting phase resulting from holon condensation we found
that the zero mode gives rise to a mass to the Dirac
fermion\cite{Kim_SU2}. As a consequence the massive spinons are
deconfined in the underdoped superconducting state. We call this
spin state a gapped $U1SL$. This deconfinement mechanism is
totally different from that in the undoped $U1SL$. In the $U1SL$
critical fluctuations of the Dirac spinons are expected to
suppress the instanton excitations and thus the massless spinons
are deconfined\cite{U1SL,Senthil_Lee}. We find that the massive
spinon in the gapped $U1SL$ is localized in the non-magnetic
disorder in contrast with the case of the gapless $U1SL$. In the
critical $U1SL$ the massless spinon remains delocalized even in
the presence of the weak disorder\cite{Kim_U1SL} as discussed
above. Owing to the localized spin the non-magnetic disorder acts
as magnetic one.

First we review a one dimensional hole doped Mott insulator.
Reviewing this, we can understand a role of doped holes in a Mott
insulator. Further, we will see that doped holes in a two
dimensional doped $U1SL$ play a similar role as the case of the
one dimensional doped Mott insulator. We consider the $t-J$
Hamiltonian to describe a doped Mott insulator \bqa && H =
-t\sum_{i=1}^{N}(c_{\sigma{i}}^{\dagger}c_{\sigma{i+1}} + h.c.) +
J\sum_{i=1}^{N}{\bf S}_{i}\cdot{\bf S}_{i+1} . \eqa  In the
absence of hole doping hopping of electrons denoted by the first
term is suppressed and thus the $t-J$ Hamiltonian is reduced to
the Heisenberg Hamiltonian describing a quantum antiferromagnetic
spin chain. Low energy physics of the quantum spin chain can be
described by a non-linear $\sigma$ model with a Berry phase
term\cite{Nagaosa}. Utilizing $CP^1$ representation, one can
represent the non-linear $\sigma$ model in terms of bosonic
spinons interacting via compact U(1) gauge fields in the presence
of the Berry phase\cite{Nagaosa}. In the case of integer spin the
Berry phase is ignorable. Strong quantum fluctuations originating
from low dimensionality lead the bosonic spinons to be
massive\cite{Nagaosa}. The integer spin chain becomes disordered.
The massive spinons are confined via a linearly increasing gauge
potential in a distance which results from Maxwell kinetic energy
of the gauge field\cite{Park_Kwon,Deconfinement}. As a consequence
mesons (spinon-antispinon bound states, here spin excitons) are
expected to appear\cite{Park_Kwon,Deconfinement}. In the case of
half-odd integer spin the Berry phase plays a crucial role to
cause destructive interference between quantum fluctuations, thus
weakening quantum fluctuations. Owing to the Berry phase
contribution the half-odd integer spin chain is expected to be
ordered. But low dimensionality leads the system not to be ordered
but to be critical\cite{Nagaosa}. As a result the spinons are
massless\cite{Park_Kwon,Deconfinement}. The massless spinons are
expected to be deconfined\cite{Deconfinement} because critical
fluctuations of the spinons can weaken gauge fluctuations via
screening.

Now we consider hole doping to the spin chain. Then hopping of
doped holes is admitted. Shankar showed that the doped holes are
represented by massless Dirac fermions and the fermionic holes
interact with the bosonic spinons via U(1) gauge
fields\cite{Shankar}. A low energy effective field theory is
obtained to be\cite{Shankar}  \bqa && S = \int{d^2x} \Bigl[
\frac{1}{2g}|(\partial_{\mu} - ia_{\mu})z_{\sigma}|^2 +
m^2|z_{\sigma}|^2 \nn && + \frac{u}{2}(|z_{\sigma}|^2)^2 -
iS\epsilon_{\mu\nu}\partial_{\mu}a_{\nu} \Bigr] \nn && +
\int{d^2x} \Bigl[
\bar{\psi}_{A}\gamma_{\mu}(\partial_{\mu}+ia_{\mu})\psi_{A} +
\bar{\psi}_{B}\gamma_{\mu}(\partial_{\mu}-ia_{\mu})\psi_{B} \Bigr]
. \eqa Here $z_{\sigma}$ is a bosonic spinon (spin) and
$\psi_{A(B)}$, a fermionic holon (charge) in a sublattice $A(B)$.
The spinons and holons interact via the compact U(1) gauge field
$a_{\mu}$. $g^{-1}$ is a stiffness parameter, $m^2$, a mass of the
spinons, and $u$, a strength of
self-interaction\cite{Deconfinement}. A detailed derivation is
given by Ref. \cite{Shankar}. A key question is what an effect of
the massless Dirac fermions on spinon dynamics is. The massless
Dirac fermions are shown to kill the Berry phase
effect\cite{Witten}. In order to see this we utilize a standard
bosonization method\cite{Witten,Shankar} \bqa &&
\bar{\psi}_{A}\gamma_{\mu}\partial_{\mu}\psi_{A} =
\frac{1}{2}(\partial_{\mu}\phi_{A})^2 \mbox{,  }  \mbox{   }
\bar{\psi}_{B}\gamma_{\mu}\partial_{\mu}\psi_{B} =
\frac{1}{2}(\partial_{\mu}\phi_{B})^2 , \nn &&
\bar{\psi}_{A}\gamma_{\mu}\psi_{A} =
\frac{1}{\sqrt{\pi}}\epsilon_{\mu\nu}\partial_{\nu}\phi_{A}
\mbox{,  }  \mbox{   } \bar{\psi}_{B}\gamma_{\mu}\psi_{B} =
\frac{1}{\sqrt{\pi}}\epsilon_{\mu\nu}\partial_{\nu}\phi_{B} . \eqa
Here $\phi_{A}$ and $\phi_{B}$ are bosonic fields in each
sublattice. Inserting these into the above action Eq. (2), we
obtain \bqa && S = \int{d^2x} \Bigl[ \frac{1}{2g}|(\partial_{\mu}
- ia_{\mu})z_{\sigma}|^2 + m^2|z_{\sigma}|^2 +
\frac{u}{2}(|z_{\sigma}|^2)^2 \Bigr] \nn && + \int{d^2x} \Bigl[
\frac{1}{2}(\partial_{\mu}\phi_{+})^2 +
\frac{1}{2}(\partial_{\mu}\phi_{-})^2 \nn && +
i\sqrt{\frac{2}{\pi}}\phi_{-}\epsilon_{\mu\nu}\partial_{\mu}a_{\nu}
- iS\epsilon_{\mu\nu}\partial_{\mu}a_{\nu} \Bigr] \eqa with
$\phi_{+} = \frac{1}{\sqrt{2}}(\phi_{A} + \phi_{B})$ and $\phi_{-}
= \frac{1}{\sqrt{2}}(\phi_{A} - \phi_{B})$. Shifting the
$\phi_{-}$ field to $\phi_{-} + \sqrt{\frac{\pi}{2}}S$, we can
easily see that the Berry phase term is wiped out from the action.
Thus half-odd integer spin chains are not distinguishable from
integer spin chains. The bosonic spinon in the doped half-odd
integer spin chain is expected to be massive like that in the
undoped integer spin chain. But the spinons here are not confined
in contrast with the case of integer spin
chains\cite{Witten,Shankar}. Integrating over the $\phi_{-}$
field, we find that the U(1) gauge field becomes massive and thus
it is ignorable in the low energy limit. As a consequence the
massive spinons are deconfined. A U(1) spin liquid with massive
spinon excitations emerges in a doped antiferromagnetic spin
chain. If we introduce an electromagnetic field $A_{\mu}$, we
obtain a coupling term of
$i\sqrt{\frac{2}{\pi}}\phi_{+}\epsilon_{\mu\nu}\partial_{\mu}A_{\nu}$.
Integrating over the $\phi_{+}$ field, we obtain a mass of the
electromagnetic field. This implies superconductivity in the doped
spin chain, which is consistent with the result of
Shankar\cite{Shankar}. In summary, doped holes lead a spin degree
of freedom to be a gapped $U1SL$ and a charge degree of freedom to
be superconducting. It should be noted that this result is exact
in the low energy limit\cite{Shankar}. As will be seen below,
these roles of the doped holes in a one dimensional Mott insulator
are very similar to that in a two dimensional one.

Now we investigate a role of non-magnetic disorder in this massive
$U1SL$. An impurity action is given by $S_{imp} = \int{d^2x}
V(x)|z_{\sigma}|^2$\cite{Impurity_electron_coupling}. Here $V(x)$
is a random potential resulting from quenched disorder. The random
potential is random only in space but static in time, i.e., $V(x)
= V({r})$. We assume that $V({r})$ is a gaussian random potential
with $<V({r})V({r'})> = W\delta({r} - {r'})$ and $<V({r})> = 0$.
Integrating over the massive U(1) gauge field $a_{\mu}$ and using
the standard replica trick to average over the gaussian random
potential, we obtain an effective action in the presence of
non-magnetic disorder \bqa && S_{eff} = \int{dr}{d\tau}
\sum_{\alpha=1}^{N}\Bigl[
\frac{1}{2g}|\partial_{\mu}z_{\sigma\alpha}|^2 +
m^2|z_{\sigma\alpha}|^2 + \frac{u}{2}(|z_{\sigma\alpha}|^2)^2
\Bigr] \nn && -
\frac{W}{2}\int{dr}{d\tau_1}{d\tau_2}\sum_{\alpha,\beta=1}^{N}|z_{\sigma\alpha}(\tau_1)|^2|z_{\sigma\beta}(\tau_2)|^2
. \eqa Here $\alpha, \beta$ are replica indices and the limit $N
\rightarrow 0$ is to be taken at the end. In the above a local
current interaction originating from integration over the massive
gauge field $a_{\mu}$ is not explicitly taken into account because
it is expected to be irrelevant (marginally) in the low energy
limit. From this effective action one can easily see a fate of the
spinon. It is well known that the non-magnetic disorder results in
localization of massive particles in
$(1+1)D$\cite{Kim_U1SL,Impurity_1}. The spinon is trapped by the
disorder. The non-magnetic disorder is expected to act as magnetic
disorder owing to the trapped spinon.

Next we consider a two dimensional doped $U1SL$ which is suggested
to describe physics of the pseudogap phase\cite{Senthil_Lee,U1SL}.
The problem of hole doping to the $U1SL$ is examined in the
context of a SU(2) slave boson theory developed by Lee, Wen, and
coworkers\cite{SU2_gauge}. Following Wen and Lee, we consider the
staggered flux phase as an ansatz for the pseudogap state. In the
staggered flux phase a spin degree of freedom is described by
$QED_3$ in terms of the massless Dirac spinons interacting via
compact U(1) gauge fields. Doped holes are represented by holons
carrying only a charge degree of freedom\cite{Slave_boson}. In the
SU(2) slave boson theory the holons have two components in
association with the SU(2) symmetry\cite{SU2_symmetry}. An
effective Lagrangian for the holon field is given by a non-linear
$\sigma$ model coupled to the spinons via the compact U(1) gauge
field\cite{SU2_gauge}. The problem of hole doing to the $U1SL$ is
investigated by an effective Lagrangian in the staggered flux
phase\cite{Kim_SU2} \bqa && Z =
\int{D\psi_\alpha}{Dz_\beta}{Da^{3}_{\mu}}e^{-\int{d^3x} {\cal L}}
, \nn && {\cal L} =
\bar{\psi}_{\alpha}\gamma_{\mu}(\partial_{\mu}\delta_{\alpha\beta}
- \frac{i}{2}a^{3}_{\mu}\tau^{3}_{\alpha\beta})\psi_{\beta} \nn &&
+ \frac{1}{2g}|(\partial_{\mu}\delta_{\alpha\beta} -
\frac{i}{2}a^{3}_{\mu}\tau^{3}_{\alpha\beta} )z_{\beta}|^2 +
V(z_{\alpha}) \nn && +
G\bar{\psi}_{\alpha}\tau^{k}_{\alpha\beta}\psi_{\beta}z^{\dagger}_{\gamma}\tau^{k}_{\gamma\delta}z_{\delta}
+ \frac{1}{2e_{in}^2}|\partial\times{a^3}|^2 . \eqa Here
$\psi_{\alpha}$ is a $4$ component massless Dirac fermion with an
isospin index $\alpha = 1, 2$, and $z_{\alpha}$, a phase field of
a holon doublet\cite{SU2_symmetry}. $V(z_{\alpha})$ is an
effective potential for easy plane anisotropy, resulting from
contributions of high energy fermions\cite{Kim_SU2,SU2_gauge}.
$g^{-1} \sim x$ is a phase stiffness of the holon field with hole
concentration $x$, and $G \sim x$, a coupling constant between
spinon and holon isospins. $\tau^k$ acts on SU(2) isospin space.
$e_{in}$ is an effective internal gauge charge. The spinons and
holons interact via not only the gauge field $a_{\mu}^{3}$ but
also their isospins. The coupling between the spinon and holon
isospins is expected to result from gauge interactions mediated by
the time component of the SU(2) gauge fields\cite{Kim_SU2}.
Similar consideration can be found in Ref. \cite{SU2_gauge}. The
kinetic energy of the gauge field results from particle-hole
excitations of high energy quasiparticles. Utilizing this
effective Lagrangian, recently we showed that antiferromagnetism
($AFM$) can coexist with $d-wave$ superconductivity
($dSC$)\cite{Kim_SU2}. The $AFM$ results from the Dirac fermions
and the $dSC$, holon condensation. Holon condensation is shown to
result in a zero mode of a nodal fermion (spinon) in a single
instanton potential\cite{Kim_SU2}. Thus instanton excitations are
suppressed. The coupling between the spinon and holon isospins is
very crucial for existence of the zero mode. If the term is
ignored, the fermion zero mode is not found\cite{Marston}. Since
the coupling constant $G$ is proportional to hole concentration
$x$, there exists no fermion zero mode at half filling.
Suppression of instanton excitations resulting from the fermion
zero mode leads to deconfinement of the spinons and holons at $T =
0$ K in underdoped superconducting phase. Further, an instanton in
the presence of the fermion zero mode induces a fermion
mass\cite{Kim_SU2}. The mass results in
$AFM$\cite{Don_Kim,Marston}. Thus the $AFM$ of the nodal fermions
is expected to coexist with the $dSC$ in the underdoped cuprates.
Despite the coexistence the $AFM$ has little effect on the
$dSC$\cite{Kim_SU2}. As a result a superconductor to insulator
transition in the underdoped region is found to fall in the XY
universality class consistent with experiments\cite{Kim_PRL}.

Now we derive an effective action for the massive Dirac fermions
as we do Eq. (5) in $(1+1)D$. Performing a standard duality
transformation of the non-linear $\sigma$ model for the holon
fields\cite{Kim_SU2,Deconfinement}, we obtain  \bqa && S =
\int{d^3x} \Bigl[ |(\partial_{\mu} -
ic_{\uparrow\mu})\Phi_{\uparrow}|^{2} + |(\partial_{\mu} -
ic_{\downarrow\mu})\Phi_{\downarrow}|^{2} \nn && +
V(|\Phi_{\uparrow}|,|\Phi_{\downarrow}| ) +
\frac{1}{2\rho}|\partial\times{c}_{\uparrow}|^2 +
\frac{1}{2\rho}|\partial\times{c}_{\downarrow}|^2 \nn && +
\frac{1}{2e_{in}^2}|\partial\times{a^3}|^2 - i
(\partial\times{a}^{3})_{\mu}(c_{\uparrow\mu} - c_{\downarrow\mu})
\nn && +
\bar{\psi}_{\alpha}\gamma_{\mu}(\partial_{\mu}\delta_{\alpha\beta}
- \frac{i}{2}a^{3}_{\mu}\tau^{3}_{\alpha\beta})\psi_{\beta} +
m_{\psi}\bar{\psi}_{\alpha}\psi_{\alpha} \Bigr] . \eqa Here
$\Phi_{\uparrow(\downarrow)}$ is a vortex field with isospin
$\uparrow (\downarrow)$ and $c_{\uparrow(\downarrow)\mu}$, its
corresponding vortex gauge field mediating interactions between
the vortices. $V(|\Phi_{\uparrow}|,|\Phi_{\downarrow}| )$ is an
effective potential including vortex mass and self-interactions.
$\rho \sim g^{-1} \sim x$ originates from the phase stiffness with
the hole concentration $x$. $m_\psi$ is a mass of the nodal
fermion. Suppression of instantons originating from the fermion
zero mode leads the U(1) gauge field $a_{\mu}^{3}$ to be
non-compact. In the superconducting state the vortex fields are
not condensed. As a result the vortex gauge fields remain
massless. Integrating over the massless vortex gauge fields, we
obtain a massive U(1) gauge field. We note that in one dimension
integration over the $\phi_{-}$ field results in the massive gauge
field. Integrating over the massive internal gauge field, we
obtain an effective action for the massive spinons which is
essentially same as Eq. (5) (in the absence of the disorder)  \bqa
&& S_{eff} = \int{d^3x} \Bigl[
\bar{\psi}_{\alpha}\gamma_{\mu}\partial_{\mu}\psi_{\alpha} +
m_{\psi}\bar{\psi}_{\alpha}\psi_{\alpha} \Bigr]  . \eqa Here a
local current interaction term
$\frac{1}{\rho}|\bar{\psi}_{\alpha}\gamma_{\mu}\psi_{\alpha}|^2$
generated by integration over the massive gauge field is
irrelevant in the low energy limit and thus can be safely ignored.
In summary, holon condensation in the doped $U1SL$ leads a spin
degree of freedom to be a gapped $U1SL$ and a charge degree of
freedom to be superconducting. This is essentially same as the one
dimensional case except the fact that the spinon is a fermion here
while it is a boson in one dimensional case.

Now we examine a role of the non-magnetic impurities in the U(1)
spin liquid with the massive spinons. An impurity action is given
by\cite{Kim_U1SL} \bqa  S_{imp} = \int {d^3x}
V(x)\bar{\psi}_{\alpha}\gamma_{0}{\psi}_{\alpha}  . \eqa Here
$V(x)$ is a random potential as discussed in one dimension. It
does not depend on time, i.e., $V(x) = V({\bf r})$. We also assume
that $V({\bf r})$ is a gaussian random potential with $<V({\bf
r})V({\bf r'})> = W\delta({\bf r} - {\bf r'})$ and $<V({\bf r})> =
0$. Using the standard replica trick to average over the gaussian
random potential, we obtain an effective action in the presence of
the non-magnetic disorder \bqa && S_{eff} = \int{d^2{\bf
r}}{d\tau} \sum_{k = 1}^{M} \Bigl[
\bar{\psi}_{\alpha{k}}\gamma_{\mu}\partial_{\mu}\psi_{\alpha{k}} +
m_{\psi}\bar{\psi}_{\alpha{k}}\psi_{\alpha{k}} \Bigr] \nn && -
\frac{W}{2}\int{d^2{\bf
r}}{d\tau_1}{d\tau_2}\sum_{k,l=1}^{M}(\bar{\psi}_{\alpha{k}}\gamma_{0}{\psi}_{\alpha{k}})(\tau_1)
(\bar{\psi}_{\alpha{l}}\gamma_{0}{\psi}_{\alpha{l}})(\tau_2) .
\eqa Here $k, l$ are replica indices and the limit $M \rightarrow
0$ is to be taken at the end. This effective action is essentially
same as Eq. [5] except the fact that the spinons are fermions
here. It is well known that the massive Dirac fermion is localized
by the non-magnetic impurities in $(2+1)D$\cite{Kim_U1SL}. The
non-magnetic impurity is expected to act as a magnetic impurity.

We believe that the localized magnetic moment is not screened by
other spinons. This is because the spinons are massive and thus
there are no density of states at the Fermi energy. As a result
the localized magnetic moment is expected to act as a free spin.
We think that this is the origin to cause a Curie-Weiss behavior
of a magnetic susceptibility\cite{Theory_1,Theory_2,Theory_4} and
enhancement of antiferromagnetic correlations\cite{Theory_1_1}.
But we admit that our conclusion is based on $T = 0$ K
calculation. It is not clear whether our results hold at $T \not=
0$ K. This is because an instanton effect is not clearly
understood at finite temperature. The role of non-magnetic
disorder in a doped $U1SL$ at finite temperature is left for
future study.

Hole doping to a Mott insulator leads a critical spin liquid to a
gapped spin liquid in both one and two dimensions. We examined the
role of non-magnetic impurities in the gapped spin liquid. We
showed that the massive spinon is localized in the non-magnetic
disorder. The localized magnetic moment in the non-magnetic
disorder is expected to explain the Curie-Weiss behavior of a
magnetic susceptibility\cite{Theory_1,Theory_2,Theory_4}.

K.-S. Kim thanks Dr. Park, Kee-Su for introducing Ref.
\cite{Witten}.


\begin{thebibliography}{9}
\bibitem{Impurity_electron_coupling} An impurity Hamiltonian is given by \bqa H_{imp} =
\sum_{i=1}^{N}V_{i}c_{\sigma{i}}^{\dagger}c_{\sigma{i}} \nonumber
. \eqa Using the slave-fermion constraint
$\sum_{\sigma=1}^{2}|z_{\sigma{i}}|^2 + f_{i}^{\dagger}f_{i} = 1$
with a fermionic holon $f_{i}$ at site $i$, we find that the
electron density is given by the spinon density, i.e.,
$c_{\sigma{i}}^{\dagger}c_{\sigma{i}} = |z_{\sigma{i}}|^2$. This
is proved by \bqa && c_{\sigma{i}}^{\dagger}c_{\sigma{i}} =
z_{\sigma{i}}^{\dagger}z_{\sigma{i}}f_{i}f_{i}^{\dagger} = (1 -
f_{i}^{\dagger}f_{i})f_{i}f_{i}^{\dagger} \nn && = (1 -
f_{i}^{\dagger}f_{i})(1 - f_{i}^{\dagger}f_{i}) = 1 -
2f_{i}^{\dagger}f_{i} + f_{i}^{\dagger}f_{i}f_{i}^{\dagger}f_{i}
\nn && = 1 - 2f_{i}^{\dagger}f_{i} + f_{i}^{\dagger}f_{i} = 1 -
f_{i}^{\dagger}f_{i} = z_{\sigma{i}}^{\dagger}z_{\sigma{i}}
\nonumber . \eqa
\bibitem{Slave_boson} In the one dimensional doped Mott insulator
we used a slave-fermion representation. Here a spin degree of
freedom is represented by a bosonic spinon and a charge degree of
freedom, a fermionic holon. But in the two dimensional doped Mott
insulator we used a slave-boson representation. In this case a
spin degree of freedom is represented by a fermionic spinon and a
charge degree of freedom, a bosonic holon.
\bibitem{SU2_symmetry} In the SU(2) slave boson theory a local SU(2) symmetry originates from
the symmetry of a spin $\uparrow$ spinon and a spin $\downarrow$
anti-spinon at half filling\cite{SU2_gauge}. As a result of the
SU(2) symmetry many U(1) ground states such as a staggered flux
phase, $d-wave$ pairing state, and rotated states between the two
are exactly degenerate at half filling\cite{SU2_gauge}. In
underdoped region the U(1) ground states are expected to remain
nearly degenerate. This motivates Wen and Lee to extend the SU(2)
symmetry at half filling away from half filling\cite{SU2_gauge}.
In the SU(2) slave boson theory fluctuations between the nearly
degenerate U(1) states can be easily taken into account via
introduction of the holon doublets\cite{SU2_gauge}.
\end{thebibliography}
\end{document}